\documentclass[11pt, draftclsnofoot, onecolumn, oneside]{IEEEtran}
\usepackage{epsfig}
\usepackage{amssymb}
\usepackage[centerlast]{caption}
\usepackage{latexsym,graphicx}
\usepackage{euscript,eufrak}
\usepackage{array}
\usepackage{mathrsfs}
\usepackage{booktabs}

\usepackage{multicol}

\def\BibTeX{{\rm B\kern-.05em{\sc i\kern-.025em b}\kern-.08em
    T\kern-.1667em\lower.7ex\hbox{E}\kern-.125emX}}

\newtheorem{thm}{Theorem.}
\newtheorem{cor}{Corollary.}

\newcommand{\Ref}[1]{(\ref{#1})}
\newcommand{\Fig}[1]{Fig. \ref{#1}}
\newcommand{\Rem}[1]{{\em Remark. #1:}}
\newcommand{\Thm}[1]{{\em Theorem \ref{#1}}}

 \makeatletter
 \let\NAT@parse\undefined
 \makeatother

\begin{document}

%

\title{Underlay Cognitive Radio with Full or Partial Channel
Quality Information}
%
%
\author{
{Na Yi, Yi Ma and Rahim Tafazolli}\small\\ C.C.S.R., University of Surrey,
Guildford, UK, GU2 7XH
\thanks{This work has been performed in the framework of the ICT project
ICT-217033 WHERE, which is partly funded by the European Union.
}}
\markboth{International Journal of Navigation and Observation, September 2010} {
} \maketitle

\begin{abstract}
Underlay cognitive radios (UCRs) allow a secondary user to enter a primary user's spectrum through intelligent utilization of multiuser channel quality information (CQI) and sharing of codebook. The aim of this work is to study two-user Gaussian UCR systems by assuming the full or partial knowledge of multiuser CQI. Key contribution of this work is motivated by the fact that the full knowledge of multiuser CQI is not always available. We first establish a location-aided UCR model where the secondary user is assumed to have partial CQI about the secondary-transmitter to primary-receiver link as well as full CQI about the other links. Then, new UCR approaches are proposed and carefully analyzed in terms of the secondary user's achievable rate, denoted by $C_2$, the capacity penalty to primary user, denoted by $\Delta C_1$, and capacity outage probability. Numerical examples are provided to visually compare the performance of UCRs with full knowledge of multiuser CQI and the proposed approaches with partial knowledge of multiuser CQI. 
\end{abstract}


\section{Introduction}
Cognitive radios convey a dynamic and flexible spectrum allocation 
policy that allows a secondary user to access a primary user's 
spectrum through exploitation of advanced air-interface techniques 
and intelligent utilization of multiuser side information such as 
user activity, channel quality information (CQI), message, codebook, 
location information, ... A good tutorial about cognitive radios can be 
found in \cite{D1216}, focused on the signal-processing perspective, 
and in \cite{D1460}, focused on the information-theoretic perspective. 
One group of cognitive radios is known as the interweave paradigm,
where a secondary user can opportunistically enter temporary spectrum 
holes and white spaces existing in both licensed or unlicensed radio 
spectrum \cite{Mitola99}. Fast and reliable spectrum sensing techniques 
are the key to the success of interweave cognitive radios. The other 
group of cognitive radios includes overlay and underlay paradigms, where 
the secondary user and the primary user form a cognitive interference channel 
(e.g. \cite{Maric06}-\nocite{D1251}\cite{D1279}). Specifically, the overlay 
cognitive user is able to sense the primary user's message, and
then employs advanced coding schemes such as the Gel'fand-Pinsker code
\cite{Gelfand1980} or the dirty-paper code \cite{Costa1983} for interference 
pre-cancellation. In the underlay paradigm, the secondary user enters the 
primary spectrum only when its activity will not cause considerable 
interference or capacity penalty to the primary user. Measure of interference 
requires knowledge about multiuser CQI. The focus of this paper is on 
the two-user Gaussian underlay cognitive radios (UCR).

\begin{figure}[t]
\begin{center}
\epsfig{file=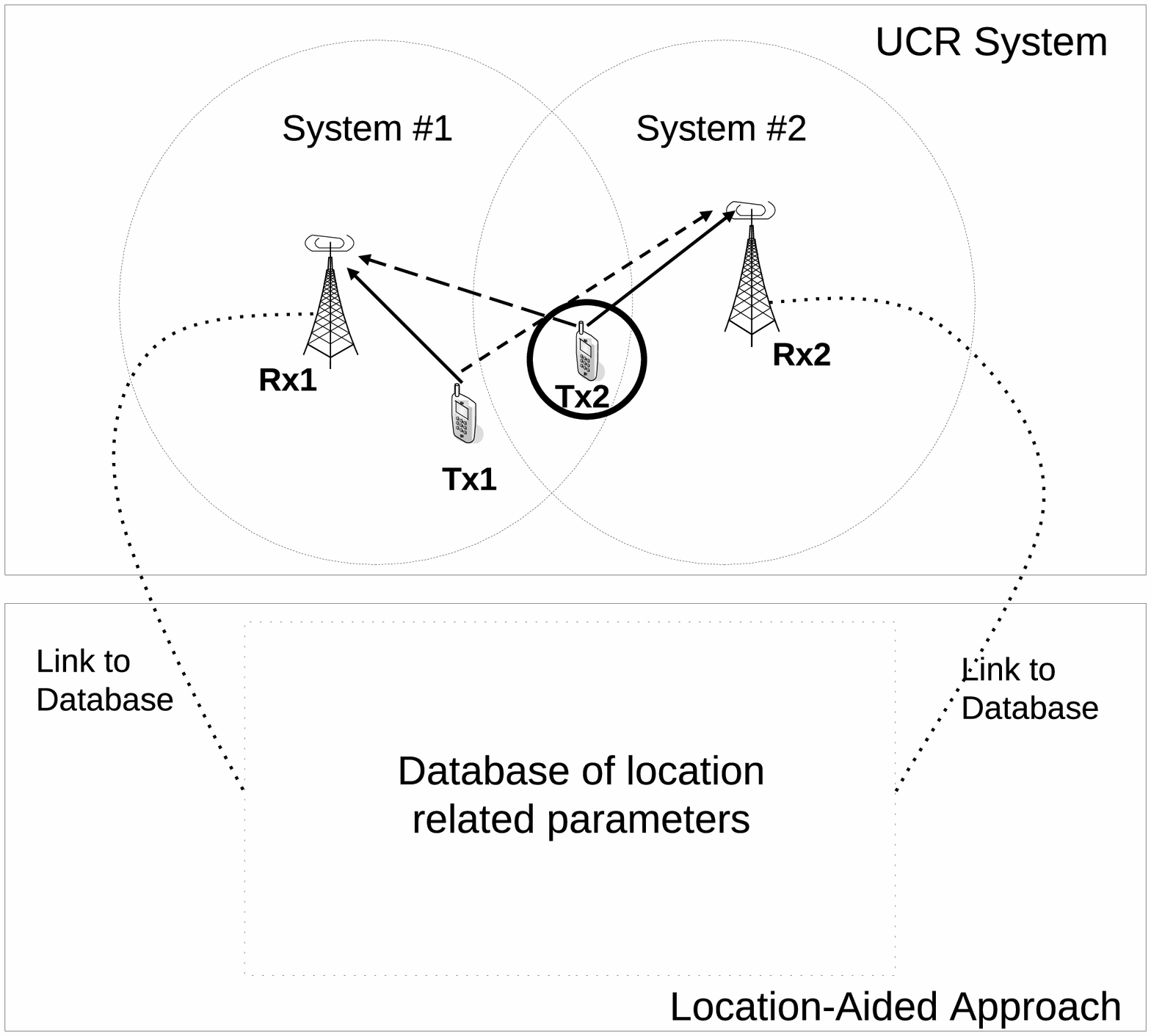, width=1\columnwidth}
\end{center}
\caption{Illustration of an example about the two-user
UCR system and a location-aided approach.}\label{figure1}
\end{figure}

\begin{figure}[t]
\begin{center}
\epsfig{file=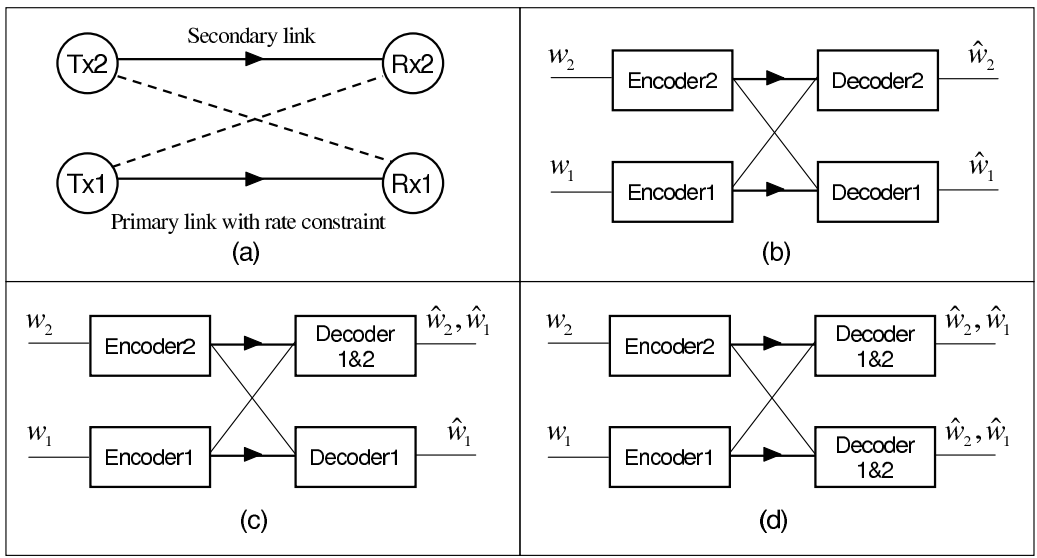, width=0.45\columnwidth, angle=270}
\end{center}
\caption{Block diagram of various UCR modes: (a) two-user
UCR channel, (b) the individual mode, (c) the CSMD mode,
(d) the TSMD mode.}\label{figure2}
\end{figure}

\Fig{figure1} illustrates an example of two-user UCR system
accommodating one primary transmitter (Tx$1$) and receiver (Rx$1$)
pair in System \#1 and one secondary transmitter (Tx$2$) and receiver (Rx$2$)
pair in System \#2. The block diagram of this UCR system is depicted in 
\Fig{figure2}-(a). In the flat-Gaussian scenario, this UCR system can be described 
as the following linear model \footnote{This is a well recognized model in the 
literature \cite{Costa1983}\nocite{Carleial78}\nocite{Sato81}\nocite{D1355}-\cite{D1637} 
where both users are assumed to employ simple random codes. 
Although rate-splitting codes have been recently introduced to cognitive radio 
channels, the focus of this paper will be on this simple system model.}. 
\begin{equation}\label{equ01}
\mathrm{Y}_1=a_{11}^{}\mathrm{X}_1+a_{21}^{}\mathrm{X}_2+\mathrm{V}_1
\end{equation}
\begin{equation}\label{equ02}
\mathrm{Y}_2=a_{12}^{}\mathrm{X}_1+a_{22}^{}\mathrm{X}_2+\mathrm{V}_2
\end{equation}
where $\mathrm{X}_i$ stands for the signal sent by the transmitter
Tx$i$ with power $P_i$ and rate $R_i$, $\mathrm{Y}_j$ for the signal
received at the receiver Rx$j$, $a_{ij}^{}$ for the channel coefficient
of the Tx$i$-Rx$j$ link, $\mathrm{V}$ for the Gaussian noise with
zero mean and variance $N_o$. This linear model shows that the UCR
system is a special case of interference channels presented in
\cite{Carleial78}-\cite{Sato81}, but the interference term
$(a_{21}^{}\mathrm{X}_2)$ in (\ref{equ01}) must not
cause considerable capacity penalty to the primary user.  According to
the multiuser-decoding capability, we can divide the UCR system into the
following four groups. Detailed introduction about these four modes
can be found in Secs. III-VI, respectively.
\subsubsection{Individual Decoding}
Both the primary user and the secondary user always deal
with the mutual interference as noise in their decoding process.

\subsubsection{Secondary-user Side Multiuser Decoding (SSMD)}
The secondary user optimally deals with the interference term
$(a_{12}^{}\mathrm{X}_1)$ in its decoding process. But, the primary user
always deals with the interference term $(a_{21}^{}\mathrm{X}_2)$ as noise.

\subsubsection{Primary-user Side Multiuser Decoding (PSMD)}
The primary user optimally deals with the interference term
$(a_{21}^{}\mathrm{X}_2)$ in its decoding process. But, the secondary user
always deals with the interference term $(a_{12}^{}\mathrm{X}_1)$ as noise.

\subsubsection{Two Sides Multiuser Decoding (TSMD)}
Both the primary user and the secondary user perform
an optimal treatment about the corresponding interference term
in their decoding process.

Key physical-layer issues this work seeks to address are mainly in
two folds:
\subsubsection*{Issue 1}
Provided full knowledge about multiuser CQI, what is the fundamental 
relationship between the secondary user's achievable rate, denoted by 
$C_2$, and capacity penalty to the primary user, denoted by $\Delta C_1$? 
What are criteria for Tx$2$ to perform efficient power allocation? Those 
questions require an answer for various UCR modes.

\subsubsection*{Issue 2}
In many practical environments, having full knowledge of CQI about all
links of the UCR system is not a suitable assumption. What are more
suitable assumptions in practice? What is the efficient UCR
strategy under new assumptions? What is the
secondary user's achievable rate? Those questions require a
satisfactory answer.

The primary objective of this work is to partially answer the above questions
through a study from the information-theoretic viewpoint. In order to focus on
the major technical issues, the following assumptions are made throughout 
this paper:
\begin{itemize}
\item[A1)] We consider a two-user UCR system accommodating one primary
transmitter-receiver pair and one secondary transmitter-receiver pair. This assumption 
can be easily assured by introducing orthogonal multiple-access schemes such as 
TDMA or FDMA to multiuser systems.
\item[A2)] Users in the system are synchronized in both the time and
frequency. Although the time-frequency synchronization is a challenge in practice, 
we would argue that the achievable rate produced under this assumption can be regarded as 
an upper bound of the practical performance.
\item[A3)] Both receivers employ maximum-likelihood (ML) detector/decoder to
offer the optimum decoding performance.
\end{itemize}
Contribution towards this work includes:
\begin{itemize}
\item[1)] The first work is to answer those questions listed in
{\em Issue 1}. Provided full knowledge of multiuser CQI, the fundamental
relationship between $C_2$ and $\Delta C_1$ is investigated for four UCR
groups. Criteria for efficient power allocation at Tx$2$ are established. The
produced results are the key to new UCR strategies proposed for the case 
with partial knowledge of the multiuser CQI.

\item[2)] As a starting point of the work towards {\em Issue 2}, we
study modeling of UCR systems in the absence of full knowledge of
multiuser CQI. After a careful justification, we establish an UCR system model,
where the secondary user is assumed to have partial knowledge of CQI
about the Tx$2$-Rx$1$ link, and have full knowledge of CQI about the
other links. Location-aided UCR is employed as an example to support our
justification.

\item[3)] We propose new spectrum-access approaches for various UCR
groups by assuming the availability of p.d.f. of CQI about the Tx$2$-Rx$1$
link. Power allocation criteria are carefully investigated in terms of $C_2$,
$\Delta C_1$, and capacity outage probability\footnote{In practice, the performance of power 
allocation will be influenced by air-interfaces and synchronization errors. The results presented 
in this paper are to provide an information-theoretic guidance to practical designs}. Assuming the channel to be
Rayleigh, numerical results are provided to visually show the performance 
of UCRs with full knowledge of multiuser CQI and the proposed approaches 
with partial knowledge of multiuser CQI. 
\end{itemize}
The rest of this paper is organized as follows. Section II offers a
brief review about capacity theorem of Gaussian interference
channel (GIC) and relate it to the UCR system. Moreover,
modeling about the UCR system with partial knowledge of multiuser CQI
is also presented. Technical contributions towards four UCR groups are
presented in Sections III-VI, respectively. Section VII draws the conclusion.

\section{System Model and Preparation}
This section first presents capacity theorem about two-user GIC and
its relationship with the UCR system, and then presents modeling of
the UCR system with partial knowledge of the multiuser CQI.

\subsection{Two-User UCR with Full Multiuser CQI}
The UCR system is a special case of interference channels. The
information-theoretic research towards interference channels started
from Carleial's work published in \cite{Carleial78}. Although
lots of research efforts have been paid in the last 30 years, capacity
region of interference channels has been found only for the case of
strong interference \cite{D1491}. To the best of our knowledge, the
state-of-the-art capacity bound of two-user GIC has recently been
reported in \cite{D1355}-\cite{D1637}. Next, we provide a brief review
about capacity theorem of two-user GIC, which offers the theoretical support
to our further investigation about the two-user UCR system.

\subsubsection{Two-User GIC with Strong Interference}
In the linear model \Ref{equ01}-\Ref{equ02}, the case of strong interference
denotes the scenario $|a_{12}^{}|\geq|a_{11}^{}|$ and $|a_{21}^{}|\geq|a_{22}^{}|$
\cite{Sato81}. In this case, the two-user GIC is in fact a compound
Gaussian multiple-access channel (MAC), whose capacity region is known
as the following union \cite{Cover06}
\begin{equation}\label{equ03}
\bigcup\left(\begin{array}{l}
R_1<\mathcal{C}\left[\gamma_{11}^{}\right],~
R_2<\mathcal{C}\left[\gamma_{22}^{}\right]
\\R_1+R_2<\min\left(\mathcal{C}\left[\gamma_{21}^{}+\gamma_{11}^{}\right],
\mathcal{C}\left[\gamma_{12}^{}+\gamma_{22}^{}\right]\right)
\end{array}
\right)
\end{equation}
where $\gamma_{ij}^{}\triangleq(P_i|a_{ij}^{}|^2)/(N_o)$ denotes the
instantaneous signal-to-noise ratio (SNR) and
$\mathcal{C}[x]\triangleq\log_2(1+x)$. Provided \Ref{equ03} and
the assumption that users share their codebook, each receiver
can reliably recover the message sent by Tx$1$ and Tx$2$, respectively.

\subsubsection{Two-User GIC with Weak or Mixed Interference}
This scenario includes cases other than the case of strong interference.
The closed-form of capacity region is unknown to this date. A look-up
table (but incomplete) about the channel capacity with respect to
various channel conditions has been reported in \cite{D1637}. Alternatively,
we can divide the two-user GIC into the following three groups with respect
to the way of dealing with the interference. The following result is
adequate for us to investigate the two-user Gaussian UCR system.

{\em Group I:} Each receiver can reliably decode the message sent by
Tx$1$ and Tx$2$, respectively. The achievable-rate region, denoted by
$\mathscr{R}^{\mathbf{I}}$, is (\ref{equ03}).

{\em Group II:} Each receiver can only decode the message sent by its
corresponding transmitter. The interference will be regarded as noise. The
achievable-rate region, denoted by $\mathscr{R}^{\mathbf{II}}$, is (see \cite{Carleial78})
\begin{equation}\label{equ04}
\mathscr{R}^{\mathbf{II}}=\bigcup\left(
\begin{array}{l}
R_1<\mathcal{C}\left[\frac{\gamma_{11}^{}}{\gamma_{21}^{}+1}\right]
\\
R_2<\mathcal{C}\left[\frac{\gamma_{22}^{}}{\gamma_{12}^{}+1}\right]
\end{array}\right).
\end{equation}

{\em Group III:} One receiver can decode the message sent by both 
transmitters, and the other can only decode the message sent by its 
corresponding transmitter. In this group, the achievable-rate region, denoted 
by $\mathscr{R}^{\mathbf{III}}$, is (see \cite{D1637})
\begin{equation}\label{equ05}
\mathscr{R}^{\mathbf{III}}=\bigcup\left(
\begin{array}{l}
R_j<\mathcal{C}\left[\frac{\gamma_{jj}^{}}{\gamma_{ij}^{}+1}\right]
\\
R_i<\mathcal{C}\left[\gamma_{ii}^{}+\gamma_{ji}^{}\right]-R_j
\\
R_i<\mathcal{C}\left[\gamma_{ii}^{}\right]
\end{array}
\right),~i\neq j.
\end{equation}
Provided $\gamma_{ij}^{}$, $_{i, j=1,2}$, we can obtain the maximum
sum-rate, $\max (R_1+R_2)$, through a comparison between
$\mathscr{R}^{\mathbf{I}}$, $\mathscr{R}^{\mathbf{II}}$, and
$\mathscr{R}^{\mathbf{III}}$.

\subsubsection{Two-User Gaussian UCR}
The UCR system is modeled as an interference channel where the primary
user wants to keep its interference-free capacity. But in many cases, the
secondary user will cause capacity penalty $\Delta C_1$ to the primary user.
Hence, the primary user's capacity is expressible as \cite{D1251}
\begin{equation}\label{equ06}
C_1=\mathcal{C}\left[\gamma_{11}^{}\right]-\Delta C_1
\end{equation}
and the secondary user's achievable rate is (see \cite{D1637})
\begin{equation}\label{equ07}
C_2=\max(R_1+R_2)-C_1.
\end{equation}
Define
\begin{equation}\label{equ08}
\Delta C_1\triangleq\rho\mathcal{C}\left[\gamma_{11}^{}\right],
\end{equation}
where $\rho$ is a positive coefficient. Eqn. (\ref{equ06}) is expressible as
\begin{equation}\label{equ09}
C_1=(1-\rho)\mathcal{C}\left[\gamma_{11}^{}\right].
\end{equation}
In order to keep the capacity penalty to be reasonably small, we usually
let $\rho\ll 1$.

\subsection{Two-User UCR with Partial Multiuser CQI}
In practice, the primary transmitter-receiver pair may operate in
the frequency-division duplex (FDD) manner, where the transmitter Tx$1$
periodically sends training sequences to support channel estimation
and coherent detection/decode at the receiver Rx$1$. Rx$1$ informs Tx$1$
regarding the CQI of Tx$1$-Rx$1$ link through a feedback channel. 
On the secondary-user side, we assume that
Rx$2$ can communicate with Tx$2$ through a feedback channel. This feedback 
channel is orthogonal to the primary user's frequency band and mainly for 
the purpose of signaling. Based on the above
system description, we provide the following justification of assumptions
about the knowledge of CQI:
\begin{itemize}
\item The secondary receiver Rx$2$ listens the conversation between
Tx$1$ and Rx$1$. Then, Rx$2$ can estimate the CQI of Tx$1$-Rx$2$
link.

\item We assume that Rx$1$ employs a simple common codebook such
as repetition code to perform the feedback of CQI. Then, Rx$2$ can
obtain the CQI about Tx$1$-Rx$1$ link through sensing of the primary user's
feedback channel.

\item At the beginning of cognitive communication, Rx$2$ requests Tx$2$
to send a training sequence over the primary spectrum. This offers the
knowledge of CQI about the Tx$2$-Rx$2$ link, but introduces
a short burst of interference to the primary user. We argue that this 
burst of interference will not cause considerable performance loss to 
the primary user.

\item Rx$1$ may estimate the CQI of Tx$2$-Rx$1$ link if appropriate,
but does not show this information in its feedback channel due to
an upper-layer protocol. In this case, Rx$2$ cannot know the CQI of
Tx$2$-Rx$1$ link. Then, our assumption is that the secondary user
knows the p.d.f. of CQI about the Tx$2$-Rx$1$ link. This assumption
is suitable for a scenario such as where the secondary user has the location
information about itself and the primary user. The secondary user can
access a well-designed and maintained database, which records the p.d.f.
of CQI between two locations. Fig. \ref{figure1} illustrates an example of
location-aided UCR system, where Rx$1$ and Rx$2$ are fixed network
nodes such as base-stations or access points, and Tx$1$ and Tx$2$ are
mobile stations. The database has a look-up table about the p.d.f. of CQI
between each fixed network node and a certain area such as the black
circle with solid line. Provided the location of Tx$2$, Rx$2$ knows which
circle Tx$2$ is currently in, and thus can look up the database to find out 
the p.d.f. of CQI about the Tx$2$-Rx$1$ link. Recently, how to design and 
maintain the location-related database is becoming an important research topic.
However, it is out of the scope of this paper. Further information about
location estimation and location-related database can be found in
European ICT WHERE \cite{WHERE01}.
\end{itemize}
As a summary, when we investigate the UCR strategy with partial
multiuser CQI, the following assumptions are made in addition to A1-A3:
\begin{itemize}
\item[A4)] Rx$2$ has full knowledge of CQI about the Tx$1$-Rx$1$ link,
the Tx$1$-Rx$2$ link, and the Tx$2$-Rx$2$ link, but only knows p.d.f.
of CQI about the Tx$2$-Rx$1$ link, denoted by $p(|a_{21}^{}|^2)$, as
well as the mean $E(|a_{21}|^2)$.

\item[A5)] Rx$2$ determines the secondary user's power and
transmission rate, and then informs Tx$2$ through the feedback channel.
\end{itemize}

\section{The Individual Decoding Mode}
\Fig{figure2}-(b) depicts the individual decoding mode where each receiver
only wants to decode the message sent by its corresponding transmitter,
and deals with the corresponding interference as noise. This mode is suitable
for the following cognitive radio scenarios:
\begin{itemize}
\item Both the primary user and secondary user employ their private
codebook;
\item Even if both users employ a common codebook, each receiver
cannot decode the other user's message due to reasons such
as channel conditions and upper layer protocols, etc.
\end{itemize}
In this situation, the UCR system can be regarded as a simple collection
of individual links. This mode has recently received an intensive
investigation in both the basic and system-level research, e.g. in
\cite{Xing2007}-\cite{Hoang2006}.

\subsection{Capacity Results with Full Knowledge of CQI}
This simple mode is already mature in terms of capacity results. The
channel capacity for both users is given by \Ref{equ04}. The capacity
penalty $\Delta C_1$ is calculated as
\begin{equation}\label{equ10}
\Delta C_1=\mathcal{C}[\gamma_{11}^{}]-\mathcal{C}\left[\frac{\gamma_{11}^{}}
{\gamma_{21}^{}+1}\right].
\end{equation}
Eqns. \Ref{equ04} and \Ref{equ10} show a known result that increasing
the secondary user's power $P_2$ will increase both $C_2$ and $\Delta C_1$.
Applying \Ref{equ08} into \Ref{equ10}, we can relate $P_2$ to the
capacity-penalty coefficient $\rho$ as
\begin{equation}\label{equ11}
P_2=\frac{1}{|a_{21}^{}|^2}\left(\frac{\gamma_{11}^{}}
{(1+\gamma_{11}^{})^{(1-\rho)}-1}-1\right).
\end{equation}
Given a coefficient $\rho$, the secondary user's power should be no larger
than \Ref{equ11}. Otherwise, the primary user would suffer capacity outage.

\Rem{1} A remarkable issue is that $\Delta C_1$ in \Ref{equ10} is a
monotonically decreasing function of $\gamma_{11}^{}$ due to the partial
derivative $(\partial \Delta C_1)/(\partial\gamma_{11}^{})<0$. This means
that the primary user operating at a high-SNR scenario is
less sensitive to the interference. Considering a high-SNR scenario that
fulfills the conditions C1) $\gamma_{11}^{}\gg 1$,
and C2) $\gamma_{11}^{}\gg\gamma_{21}^{}$,
\Ref{equ10} approximates to
\begin{equation}\label{equ12}
\Delta C_1\approx\mathcal{C}\left[\gamma_{21}^{}\right].
\end{equation}
Plugging \Ref{equ08} into \Ref{equ12} leads to
\begin{equation}\label{equ13}
P_2\approx\frac{N_o}{|a_{21}|^2}\left(\left(1+\gamma_{11}^{}\right)^\rho
-1\right).
\end{equation}
This simplified result can be utilized to allocate the secondary user's power
when the primary user operates in the high-SNR range.

\subsection{The UCR Strategy with Partial Multiuser CQI}
Section III-A shows that, provided the power $P_2$, the secondary user can
employ \Ref{equ04} to configure its transmission rate. However,
using \Ref{equ11} or \Ref{equ13} to configure $P_2$ requires the knowledge
about $|a_{21}^{}|^2$, which supposes to be unknown in some situations.
Next, we propose a new power-allocation criterion based on the
assumption A4).

{\em Criterion 1.} The power $P_2$ should be appropriately configured
so that the capacity-outage probability of primary user is not larger than
a given threshold $\mathcal{O}_\mathrm{t}$.

Based on {\em Criterion 1}, the power-allocation strategy can be
summarized into the following two steps:

{\em Step 1:} Outage probability to the primary user is a function of
the SNR mean of Tx$2$-Rx$1$ link denoted by $\bar{\gamma}_{21}^{}=(P_2E(|a_{21}^{}|^2))/(N_o)$. Motivated by this fact,
the secondary user can first calculate the outage probability for a given
$p(|a_{21}^{}|^2)$, and then determine a threshold $\bar{\gamma}_\mathrm{t}$
corresponding to $\mathcal{O}_\mathrm{t}$.

{\em Step 2:} The secondary user can access the primary spectrum for
the condition $\bar{\gamma}_{21}^{}\leq\bar{\gamma}_\mathrm{t}$. The maximum
of $P_2$ is therefore given by
\begin{equation}\label{equ14}
\max(P_2)=\frac{\bar{\gamma}_\mathrm{t}N_o}{E(|a_{21}^{}|^2)}.
\end{equation}
The secondary user's transmission rate can be calculated by applying
\Ref{equ14} into \Ref{equ04}. Next, we will use a numerical example to
introduce about how to determine the threshold $\bar{\gamma}_\mathrm{t}$,
and to show the performance in this example.

\subsection{Numerical Example}
Define an instantaneous SNR threshold $\gamma_\mathrm{t}$ as
\begin{equation}\label{equ15}
\gamma_\mathrm{t}\triangleq
\frac{\gamma_{11}^{}}{(1+\gamma_{11}^{})^{(1-\rho)}-1}-1.
\end{equation}
\Ref{equ11} indicates that the secondary user will cause capacity outage
to the primary user when $\gamma_{21}>\gamma_\mathrm{t}$. Assume the
p.d.f. $p(|a_{21}^{}|)$ to be Rayleigh as an example. The probability for the event
$(\gamma_{21}>\gamma_\mathrm{t})$ to happen can be calculated as \cite{Simon2005}
\begin{equation}\label{equ16}
\mathrm{Pr}(\gamma_{21}^{}>\gamma_\mathrm{t})=\exp\left(-\frac{\gamma_\mathrm{t}}
{\bar{\gamma}_{21}^{}}\right)\leq\exp\left(-\frac{\gamma_\mathrm{t}}
{\bar{\gamma}_\mathrm{t}}\right)
\end{equation}
where $\mathrm{Pr}(\cdot)$ denotes the probability. According to {\em Criterion 1},
the threshold $\bar{\gamma}_\mathrm{t}$ should be carefully chosen to fulfill
the following condition
\begin{equation}\label{equ17}
\exp\left(-\frac{\gamma_\mathrm{t}}{\bar{\gamma}_\mathrm{t}}\right)\leq \mathcal{O}_\mathrm{t}.
\end{equation}
We apply the definition of $\gamma_\mathrm{t}$ \Ref{equ15} in \Ref{equ17}
and obtain
\begin{equation}\label{equ18}
\bar{\gamma}_\mathrm{t}=\frac{(1+\gamma_{11}^{})^{(1-\rho)}-1-\gamma_{11}^{}}
{\left((1+\gamma_{11}^{})^{(1-\rho)}-1\right)\ln (\mathcal{O}_\mathrm{t})}.
\end{equation}
Moreover, when the primary user fulfills the high-SNR condition C1)-C2),
we can use \Ref{equ13} to define
\begin{equation}\label{equ19}
\gamma_\mathrm{t}\triangleq\left(1+\gamma_{11}^{}\right)^\rho-1.
\end{equation}
Applying \Ref{equ19} into \Ref{equ17} results in
\begin{equation}\label{equ20}
\bar{\gamma}_\mathrm{t}=\frac{1-(1+\gamma_{11})^\rho}
{\ln(\mathcal{O}_\mathrm{t})}.
\end{equation}

\begin{figure}[t]
\begin{center}
\epsfig{file=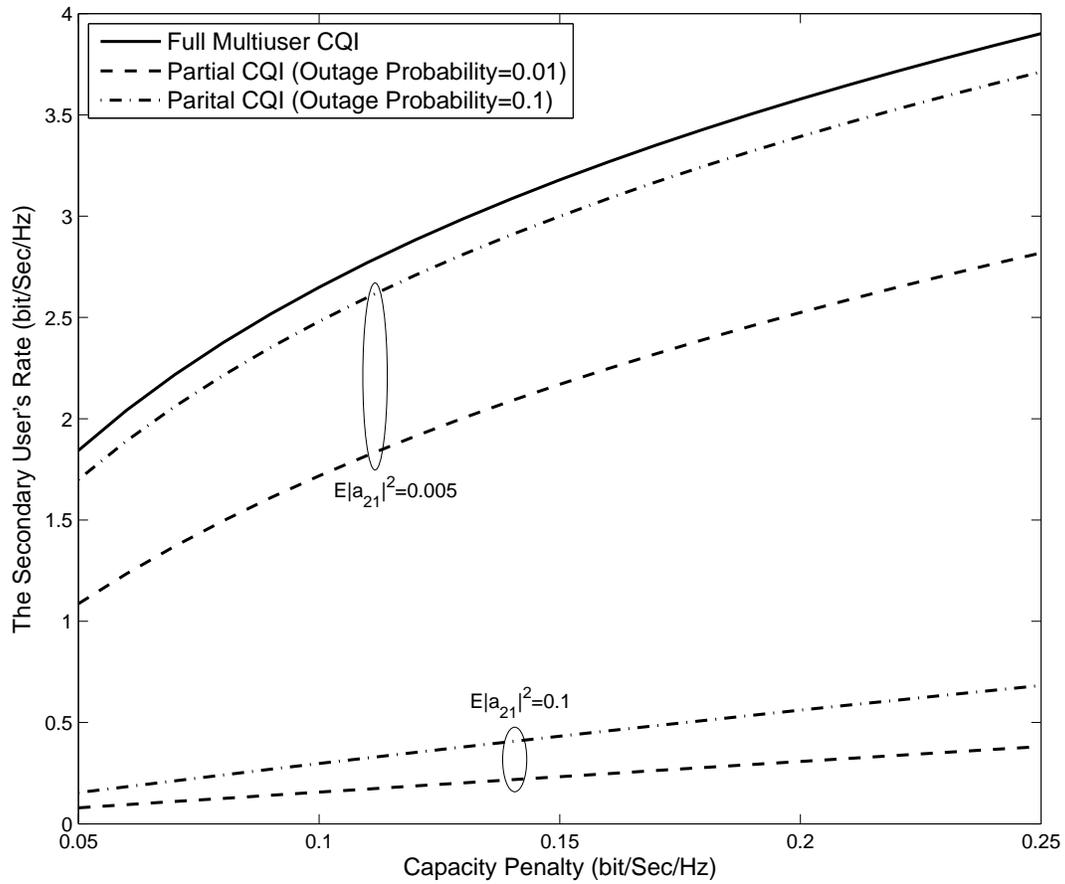, width=1\columnwidth}
\end{center}
\caption{An example of capacity results for the individual decoding mode.
The secondary user's achievable rate Vs capacity penalty to the primary
user.}
\label{figure3}
\end{figure}

\begin{figure}[t]
\begin{center}
\epsfig{file=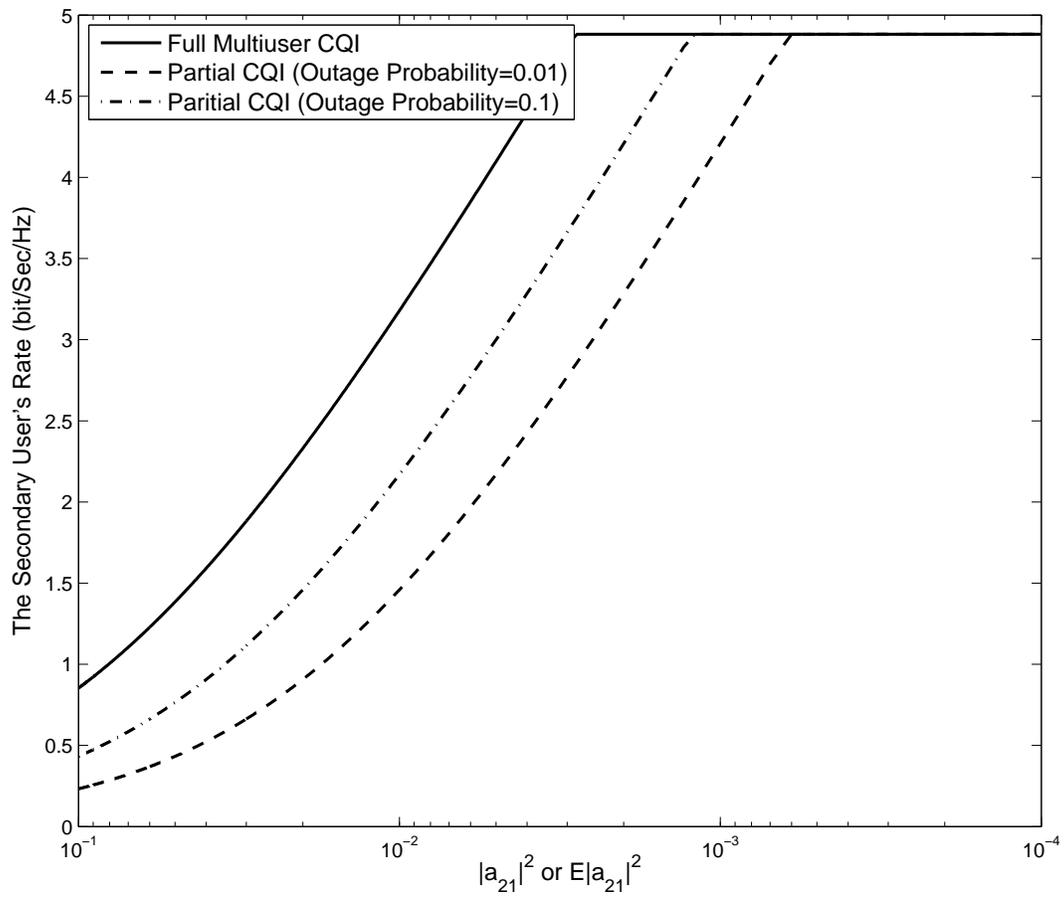, width=1\columnwidth}
\end{center}
\caption{An example of capacity results for the individual decoding mode.
The secondary user's achievable rate Vs the channel quality of
Tx$2$-Rx$1$ link.}
\label{figure4}
\end{figure}

Based on the above analytical results, we use a visual example
to exhibit the performance. In this example, the UCR system is configured
as: $|a_{11}^{}|=1$, $|a_{22}^{}|=1$, $|a_{12}^{}|=0.1$. The primary user's
power-to-noise ratio is $P_1/N_o=16$ dB. The secondary user's
power-to-noise ratio is also limited by $16$ dB. This ratio is one of
typical configurations for high-data-rate systems. For the scenario with full
multiuser CQI, we set $|a_{21}^{}|=0.1$. \Fig{figure3} illustrates the
secondary user's achievable rate (see \Ref{equ04}) against the capacity penalty
$\Delta C_1$ (see \Ref{equ10}) for cases with full or partial multiuser CQI.
It is observed that the secondary user's achievable rate generally
increases with the pay of capacity penalty to the primary user.
Moreover, in the scenario with partial multiuser CQI, the secondary
user shows increased achievable rate for the case of
larger outage probability, e.g. $\mathcal{O}_\mathrm{t}=0.1$ or
smaller $E(|a_{21}^{}|^2)$, e.g. $E(|a_{21}^{}|^2=0.005)$. \Fig{figure4}
illustrates the secondary user's achievable rate with respect to the
channel quality of Tx$2$-Rx$1$ link ($\Delta C_1=0.15$ bit/sec/Hz).
It shows that the UCR approach with partial CQI offers the
same performance as the UCR with full CQI when the Tx$2$-Rx$1$
channel is deep fade.

\section{The SSMD Mode}
\Fig{figure2}-(c) depicts the SSMD mode where each receiver wants
to decode the message sent by its corresponding transmitter. The
secondary receiver Rx$2$ will decode the primary user's message if
appropriate. The primary receiver Rx$1$ always deals with the interference
term $(a_{21}^{}\mathrm{X}_2)$ as noise. This mode is suitable for the
following cognitive radio scenario:
\begin{itemize}
\item The secondary user knows the primary user's codebook, and thus
has a chance to decode the primary user's message. This is possible
if the primary user is either using a common codebook or broadcasting
its own codebook to support, for example, user cooperation.
On the other hand, the primary user may be not aware of the existence
of secondary user, or does not know the secondary user's private codebook.
\end{itemize}
In this situation, the receiver Rx$2$ can reliably decode the primary user's
message only for the channel condition $|a_{12}^{}|\geq|a_{11}^{}|$, otherwise
the SSMD mode reduces to the individual decoding mode presented in
Section III\footnote{Multiuser information
theory about the interference channel shows that Rx$2$ can decode the signal
$\mathrm{X}_1$ if the rate of $\mathrm{X}_1$ is not larger than the achievable
rate of Tx$1$-Rx$2$ link. However, the UCR channel requires the rate of
$\mathrm{X}_1$ to be constrained only by the achievable rate of Tx$1$-Rx$1$ link.
In the case of weak interference, the Tx$1$-Rx$1$ link offers larger achievable
rate than the Tx$1$-Rx$2$ link. Rx$2$ cannot decode $\mathrm{X}_1$ if the
rate of $\mathrm{X}_1$ is larger than the achievable rate of Tx$1$-Rx$2$ link.}.
Therefore, the focus of SSMD mode is on the case $|a_{12}^{}|\geq|a_{11}^{}|$.

\subsection{Capacity Results with Full Multiuser CQI}
Suppose the channel condition $|a_{12}^{}|\geq|a_{11}^{}|$. The transmission rate
for both users is given in \Ref{equ05} by setting $i=2$ and $j=1$. More precisely,
the capacity penalty $\Delta C_1$ is \Ref{equ10}, and the secondary user's
achievable rate is expressible as
\begin{equation}\label{equ21}
C_2=\min\left(\mathcal{C}[\gamma_{12}^{}+\gamma_{22}^{}]
-(1-\rho)\mathcal{C}[\gamma_{11}^{}],~\mathcal{C}[\gamma_{22}^{}]\right).
\end{equation}
This result is subject to the power constraint of $P_2$ given in \Ref{equ11}.

\Rem{2} For the high-SNR conditions C1), C2) and the case
$|a_{12}|\geq |a_{11}|$, we can apply
(\ref{equ10}) and (\ref{equ12}) into (\ref{equ21}) to obtain
\begin{IEEEeqnarray}{ll}\label{equ22}
C_2&=\min\left(\mathcal{C}[\gamma_{12}^{}+\gamma_{22}^{}]
-\mathcal{C}[\gamma_{11}^{}]+\Delta C_1,~\mathcal{C}[\gamma_{22}^{}]\right)\nonumber
\\&\approx\min\left(\log_2\left(\frac{\gamma_{12}^{}+\gamma_{22}^{}}
{\gamma_{11}^{}}\right)+\mathcal{C}[\gamma_{21}^{}],
\mathcal{C}[\gamma_{22}^{}]\right)
\end{IEEEeqnarray}
and the transmit power $P_2$ is limited by (\ref{equ13}).
Next, we will use the above capacity results to investigate the
UCR strategy with partial multiuser CQI.

\subsection{The UCR Strategy with Partial Multiuser CQI}
Major difference between the SSMD mode and the individual decoding mode is that
the secondary user has improved achievable rate due to the availability of primary
user's codebook. However, on the primary user's side, there is no difference
between these two modes. The spectrum access and power allocation strategy
for the SSMD mode should also obey {\em Criterion 1} so as to fulfill the
requirement of outage probability. Therefore, the UCR strategy for SSMD
mode is the same as that for the individual mode, and the transmit-power
$P_2$ is limited by \Ref{equ14}. The secondary user's transmission rate
is restricted by the result produced by applying \Ref{equ14} in
\Ref{equ21}.

Apart from \Ref{equ21}, numerical results for the SSMD mode is the same
as those for the individual decoding mode. Moreover, \Ref{equ21} is also
a well-known result in the domain of multiuser information theory. Therefore, 
we do not provide a numerical example for this mode.

\section{The PSMD Mode}
This mode is referred to as a scenario where the secondary user does
not know the primary user's codebook, but share its own codebook
through upper-layer protocols. In this case, the primary user has
a chance to decode the secondary user's message, and thus has
the potential to cancel the interference caused by the secondary user.
On the other hand, the secondary user has to deal with the interference
term $(a_{12}^{}X_1)$ as noise.

\subsection{Capacity Results with Full Multiuser CQI}
In order to ensure reliable communication of the Tx$2$-Rx$2$ pair,
the secondary user's transmission rate is restricted by the second
formula in \Ref{equ04}. On the other hand, Section II-A Group III shows
that the primary user can reliably decode the secondary user's message
only when
\begin{equation}\label{equ23}
R_{2}\leq\mathcal{C}\left[\gamma_{21}^{}+\gamma_{11}^{}\right]
-C_1,
\end{equation}
where $C_1$ is given by \Ref{equ09}. Moreover, the primary user's
capacity should fulfill the following condition
\begin{equation}\label{equ24}
C_1\geq\mathcal{C}\left[\frac{\gamma_{11}^{}}{\gamma_{21}^{}+1}\right]
\end{equation}
where the interference term $(a_{21}X_2)$ in \Ref{equ01} is treated as noise.

\begin{thm}\label{Thm1}
Suppose $|a_{21}|\neq 0$ and $\rho=0$, the secondary user's
achievable rate is
\begin{equation}\label{equ25}
R_{2}\leq\mathcal{C}\left[\frac{\gamma_{22}^{}}{\gamma_{12}+1}\right]
\end{equation}
for the channel condition
\begin{equation}\label{equ26}
\frac{|a_{21}^{}|^2}{|a_{22}^{}|^2}>\frac{\gamma_{11}^{}+1}{\gamma_{12}^{}+1}
\triangleq\lambda_1
\end{equation}
otherwise
\begin{equation}\label{equ27}
R_{2}\leq\mathcal{C}\left[\gamma_{21}^{}+\gamma_{11}^{}\right]
-\mathcal{C}\left[\gamma_{11}^{}\right].
\end{equation}
\end{thm}
\begin{IEEEproof}
For the case of $\rho=0$, the results \Ref{equ04} and \Ref{equ23} show
that the secondary user's transmission rate should fulfill
\begin{IEEEeqnarray}{ll}\label{equ28}
R_2&\leq\min\left(\mathcal{C}\left[\frac{\gamma_{22}^{}}{\gamma_{12}^{}+1}\right],
\mathcal{C}\left[\gamma_{21}^{}+\gamma_{11}^{}
\right]-\mathcal{C}\left[\gamma_{11}^{}\right]\right)
\end{IEEEeqnarray}
otherwise, either the primary user or the secondary user cannot perform
reliable communication. Then, it is straightforward to justify that the
right-hand term in \Ref{equ25} is smaller than the right-hand term in
\Ref{equ27} only for the channel condition \Ref{equ26} to be satisfied.
This theorem is therefore proved.
\end{IEEEproof}

\Thm{Thm1} gives the secondary user's achievable rate subject to zero
capacity-penalty to the primary user. It can be observed that $R_2$ would
be almost zero if the channel gain $|a_{21}|$ is deep fade. This result is
inconsistent with the original idea of UCR which takes advantage of
the case $|a_{21}|\approx 0$. In other words, it is not wise to always
target on zero capacity-penalty to the primary user. Below provides
two criteria to handle the issue of capacity-penalty.

{\em Criterion 2:} The pay of capacity penalty offers 
improvement of the sum-rate of UCR, i.e. $\max(R_1+R_2)$.

{\em Criterion 3:} The capacity penalty is tolerable to
the primary user, e.g. $\rho\ll 1$.

\begin{thm}\label{Thm2}
Suppose the following channel condition
\begin{equation}\label{equ29}
\frac{|a_{21}^{}|^2}{|a_{22}^{}|^2}<\frac{1}{\gamma_{12}^{}+1}
\triangleq\lambda_2
\end{equation}
the secondary user's achievable rate is \Ref{equ25} subject to
the power constraint \Ref{equ11}.
\end{thm}
\begin{IEEEproof}
The result \Ref{equ23} indicates that the pay of capacity penalty
will not improve $\max(R_1+R_2)$ if the primary user wants to reliably
decode the secondary user's message. Hence, the only case to have an
improved $\max(R_1+R_2)$ is to deal with the interference term $(a_{21}^{}X_2)$
as noise, for which $R_2$ is only limited by \Ref{equ25}. Moreover, the
following inequality has to be satisfied so as to fulfill {\em Criterion 2}
\begin{IEEEeqnarray}{ll}\label{equ30}
&\mathcal{C}\left[\frac{\gamma_{11}^{}}{\gamma_{21}^{}+1}\right]+
\mathcal{C}\left[\frac{\gamma_{22}^{}}{\gamma_{12}^{}+1}\right]
>\mathcal{C}\left[\gamma_{11}^{}+\gamma_{21}^{}\right].
\end{IEEEeqnarray}
Solving (\ref{equ30}) leads to the channel condition \Ref{equ29}. In order to
fulfill {\em Criterion 3}, the transmit-power $P_2$ should be limited by
\Ref{equ11}. This theorem is therefore proved.
\end{IEEEproof}

According to {\em Theorems 1\&2}, we can conclude the following results:
\begin{itemize}
\item[1)] For the channel condition \Ref{equ26}, Tx$2$ can talk to
Rx$2$ at a rate \Ref{equ25} without causing capacity penalty
to the primary user. The transmit power $P_2$ is limited by the
secondary user's local power constraint.

\item[2)] For the channel condition \Ref{equ29}, Tx$2$ can talk to
Rx$2$ at a rate \Ref{equ25}. The transmit power $P_2$ is limited by
\Ref{equ11} to keep the capacity penalty $\Delta C_1$ under an acceptable
level.

\item[3)] For channel conditions other than \Ref{equ26} and \Ref{equ29},
Tx$2$ can talk to Rx$2$ at a rate \Ref{equ27} without causing capacity
penalty to the primary user. The transmit power $P_2$ is limited by the
secondary user's local power constraint.
\end{itemize}

\subsection{The UCR Strategy with Partial Multiuser CQI}
Section V-A shows that the spectrum access and power allocation
strategy for the PSMD mode requires the full knowledge of $|a_{21}^{}|$.
Here, we present a new UCR strategy under the assumption A4).
The main idea is summarized as follows.

Define a threshold of probability denoted by $\epsilon$. Based on
{\em Theorems 1\&2}, the secondary user will access the primary
spectrum for the following three cases:

\subsubsection*{Case 1}
Suppose
\begin{equation}\label{equ31}
\mathrm{Pr}\left(\frac{|a_{_{21}}|^2}{|a_{_{22}}|^2}>
\lambda_1\right)>\epsilon
\end{equation}
the secondary user will enter the primary spectrum
at a rate \Ref{equ25} with $P_2$ limited by its local power constraint.
In this case, the primary user does not have a capacity penalty, but
suffers capacity outage with the probability
$(1-\epsilon)$.

\subsubsection*{Case 2}
Suppose
\begin{equation}\label{equ32}
\mathrm{Pr}\left(\frac{|a_{_{21}}|^2}{|a_{_{22}}|^2}<
\lambda_2\right)>\epsilon
\end{equation}
the secondary user's transmission rate is also \Ref{equ25}.
In this case, the primary user deals with the interference as noise,
and thus has the capacity penalty \Ref{equ10}. Moreover, the
secondary user's power $P_2$ should be carefully configured
in terms of capacity penalty and outage probability to the primary user.
This issue will receive further investigation by employing a numerical
example.

\subsubsection*{Case 3}
Suppose
\begin{equation}\label{equ33}
\mathrm{Pr}\left(\lambda_2
\leq\frac{|a_{_{21}}|^2}{|a_{_{22}}|^2}\leq
\lambda_1\right)>\epsilon.
\end{equation}
{\em Theorem 1} shows that the secondary user can talk at a
rate \Ref{equ27}. Unfortunately, the secondary user does not know
$|a_{21}|$, and thus cannot straightforwardly employ \Ref{equ27} to
determine its achievable rate. In this case, we propose to use
the following formula produced by replacing the term
$\gamma_{21}^{}$ with $(P_2\mathcal{L})/(N_o)$ in \Ref{equ27}
\begin{equation}\label{equ34}
R_{2}\leq\mathcal{C}\left[\frac{P_2\mathcal{L}}{N_o}+\gamma_{11}^{}\right]
-\mathcal{C}\left[\gamma_{11}^{}\right]
\end{equation}
where $\mathcal{L}\in(|a_{22}^{}|^2\lambda_2,
|a_{22}^{}|^2\lambda_1)$ is a scaling factor. This case will be
further investigated through a numerical example.

Finally, for cases other than \Ref{equ31}-\Ref{equ33}, the secondary
user will not enter the primary spectrum.

\subsection{Numerical Example}
\Ref{equ31}-\Ref{equ33} shows that the proposed UCR strategy
is based on the statistical relationship between $|a_{21}^{}|$ and
$|a_{22}^{}|$. Considering $|a_{21}^{}|$ to be Rayleigh as a numerical
example, we investigate the performance of the proposed approach.

\subsubsection*{Case 1}
The key issue of this case is to find out the relationship between
$E(|a_{21}^{}|^2)$ and the threshold of outage probability
$\mathcal{O}_\mathrm{t}$, and then to link this relationship to
the spectrum-access strategy. The following result is derived for
this issue.
\begin{cor}
Given a threshold of the primary user's outage probability
$\mathcal{O}_\mathrm{t}$, the condition for \Ref{equ31} to be
satisfied is
\begin{equation}\label{equ35}
E(|a_{21}^{}|^2)>\frac{\lambda_{1}|a_{22}^{}|^2}
{\ln(1/(1-\mathcal{O}_\mathrm{t}))}.
\end{equation}
\end{cor}
\begin{IEEEproof}
We first rewrite \Ref{equ31} into
\begin{equation}\label{equ36}
\mathrm{Pr}\left(\gamma_{21}^{}>\gamma_{22}^{}
\lambda_1\right)>\epsilon.
\end{equation}
Using the result derived in \cite{Simon2005}, \Ref{equ36} becomes
\begin{equation}\label{equ37}
\exp\left(-\frac{\gamma_{22}^{}\lambda_{1}}
{\bar{\gamma}_{21}^{}}\right)>\epsilon.
\end{equation}
Given a threshold of outage probability $\mathcal{O}_\mathrm{t}$,
the probability $\epsilon$ should fulfil
$\epsilon\geq(1-\mathcal{O}_\mathrm{t})$. Applying this result
in \Ref{equ37}, we can
easily obtain \Ref{equ35} by solving the inequality.
\end{IEEEproof}

\subsubsection*{Case 2}
The key issue of this case is to find out the relationship between
$\bar{\gamma}_{21}^{}$ and the primary user's capacity penalty and
outage probability. The derived result is summarized as below, which
offers a criterion to configure the power $P_2$.

\begin{cor}
Given a probability $\epsilon$ and a threshold of outage probability
$\mathcal{O}_\mathrm{t}$, the condition for the secondary user to operate
in Case 2 is
\begin{equation}\label{equ38}
\bar{\gamma}_{21}^{}\leq\min\left(\frac{\lambda_2\gamma_{22}^{}}
{\ln\left(1/(1-\epsilon)\right)}, \bar{\gamma}_\mathrm{t}\right)
\end{equation}
where $\bar{\gamma}_\mathrm{t}$ is given by \Ref{equ18}.
\end{cor}
\begin{IEEEproof}
The first criterion for the secondary user to operate in Case 2 is
\Ref{equ32}. Following the derivation in \cite{Simon2005}, we can
easily justify that \Ref{equ32} is equivalent to
\begin{equation}\label{equ39}
\bar{\gamma}_{21}^{}\leq\frac{\lambda_2\gamma_{22}^{}}
{\ln\left(1/(1-\epsilon)\right)}.
\end{equation}
Moreover, provided the condition \Ref{equ32}, the primary user
will always deal with the interference as noise. The SNR-mean
$\bar{\gamma}_{21}^{}$ should fulfil the condition \Ref{equ14} to
ensure the primary user's outage probability under the threshold
$\mathcal{O}_\mathrm{t}$. Then, $\bar{\gamma}_{21}^{}$ should
simultaneously fulfil the conditions \Ref{equ39} and \Ref{equ14},
which leads to the result \Ref{equ38}.
\end{IEEEproof}

Once $\bar{\gamma}_{21}^{}$ is determined by employing \Ref{equ38},
we can calculate maximum of the secondary user's power as
$\max(P_2)=\Ref{equ38}/(E(|a_{21}^{}|^2))$.

\subsubsection*{Case 3}
This case includes three issues: 1)  to find the relationship between
$E(|a_{21}^{}|^2)$ and $\epsilon$ by solving \Ref{equ33}; 2)
to determine the scaling factor $\mathcal{L}$ in \Ref{equ34}; 3)
provided the condition \Ref{equ33}, {\em Theorem 1} shows that
the secondary user will suffer capacity outage for the case of
$|a_{21}^{}|^2>\lambda_1|a_{22}^{}|^2$. Then, we should calculate
the outage probability to the secondary user. Note that, in {\em Case 3},
the primary user does not suffer capacity outage.
\begin{cor}
Given a probability $\epsilon$, a necessary condition for \Ref{equ33} to
be satisfied is
\begin{equation}\label{equ40}
\gamma_{11}^{}<\epsilon^{-\gamma_{11}^{}}-1.
\end{equation}
\end{cor}
\begin{IEEEproof}
See Appendix.
\end{IEEEproof}
Usually, the probability $\epsilon$ is expected to be sufficiently
large, e.g. $\epsilon>90\%$. In this situation, we can use \Ref{equ40}
to obtain $\gamma_{11}^{}>15$ dB. It means a necessary condition
for {\em Case 3} to happen is that the primary user operates in
a high-SNR range. Provided the condition \Ref{equ40}, the secondary
user can employ \Ref{app1-1} to relate $E(|a_{21}^{}|^2)$ to $\epsilon$.

Using the scaling factor $\mathcal{L}$ in \Ref{equ34} will result in
capacity outage to the secondary user with the outage probability
\begin{equation}\label{equ41}
\mathrm{Pr}(\mathcal{L}\leq|a_{21}^{}|^2)=1-\exp(-(\mathcal{L})/(E(|a_{21}^{}|^2))).
\end{equation}
If this outage probability is required to be no larger than a threshold
$\mathcal{O}_1$, we can obtain
\begin{equation}\label{equ42}
\mathcal{L}\leq\ln\left(\frac{1}{1-\mathcal{O}_1}\right)E(|a_{21}^{}|^2).
\end{equation}
This is one criterion to determine $\mathcal{L}$. Moreover,
$\mathcal{L}$ is also limited by the range given in \Ref{equ34}.
Applying that range in \Ref{equ42} results in
\begin{equation}\label{equ43}
E(|a_{21}^{}|^2)\geq\frac{|a_{22}^{}|^2\lambda_2}{\ln\left(1/(1-\mathcal{O}_1)\right)}.
\end{equation}
Then, we can conclude the following result:
\begin{cor}
Given the threshold of outage probability $\mathcal{O}_1$, a necessary
condition for {\em Case 3} to happen is \Ref{equ43}.
\end{cor}

{\em Corollaries 3\&4} provide an answer to the first two issues of {\em Case 3}.
The last issue to concern is the probability
$\mathrm{Pr}(|a_{21}^{}|^2>\lambda_1|a_{22}^{}|^2)$ subject to the condition
\Ref{equ33}. The result is summarized as follows.
\begin{cor}
Provided the condition \Ref{equ33}, the probability for the event
$(|a_{21}^{}|^2>\lambda_1|a_{22}^{}|^2)$ to happen is smaller than
$(1)/(\gamma_{11}^{}+1)$.
\end{cor}
\begin{IEEEproof}
The probability for the event
$(|a_{21}^{}|^2>\lambda_1|a_{22}^{}|^2)$ to happen is given in \Ref{equ37},
which can be represented into
\begin{equation}\label{equ44}
\mathrm{Pr}\left(|a_{21}^{}|^2>\lambda_1|a_{22}^{}|^2\right)=
\exp\left(-\frac{|a_{22}^{}|^2\lambda_1}{E(|a_{21}^{}|^2)}\right).
\end{equation}
Provided the condition \Ref{equ33}, \Ref{app1-2} gives the maximum of
$E(|a_{21}^{}|^2)$. Since \Ref{equ44} is an increasing function of
$E(|a_{21}^{}|^2)$, we can apply \Ref{app1-2} into \Ref{equ44} and obtain
\begin{IEEEeqnarray}{ll}\label{equ45}
\mathrm{Pr}\left(|a_{21}^{}|^2>\lambda_1|a_{22}^{}|^2\right)
&\leq\exp\left(\frac{\ln(\lambda_2/\lambda_1)}{1-\lambda_2/\lambda_1}\right)
\\&\leq\exp\left(\frac{-\ln(\gamma_{11}^{}+1)}{1-\frac{1}{\gamma_{11}^{}+1}}\right).
\label{equ46}
\end{IEEEeqnarray}
The discussion about {\em Corollary 3} shows that $\gamma_{11}^{}\gg 1$ is
the necessary condition for {\em Case 3}. Therefore, the right-hand of \Ref{equ46}
approximates to $(1)/(\gamma_{11}^{}+1)$.
\end{IEEEproof}

According to {\em Corollaries 3-5}, we summarize {\em Case 3} as follows:
\subsubsection*{Step 1}
Utilize \Ref{equ40} and {\em Corollary 5} to verify whether $\gamma_{11}^{}$
fulfils the required condition. If true, go to Step 2;
\subsubsection*{Step 2}
Utilize \Ref{equ43} and \Ref{app1-2} to verify whether $E(|a_{21}^{}|^2)$
is in the appropriate range; If true, go to Step 3;
\subsubsection*{Step 3}
Utilize \Ref{equ42} to determine $\mathcal{L}$, and apply it in \Ref{equ34}.

\begin{figure}[t]
\begin{center}
\epsfig{file=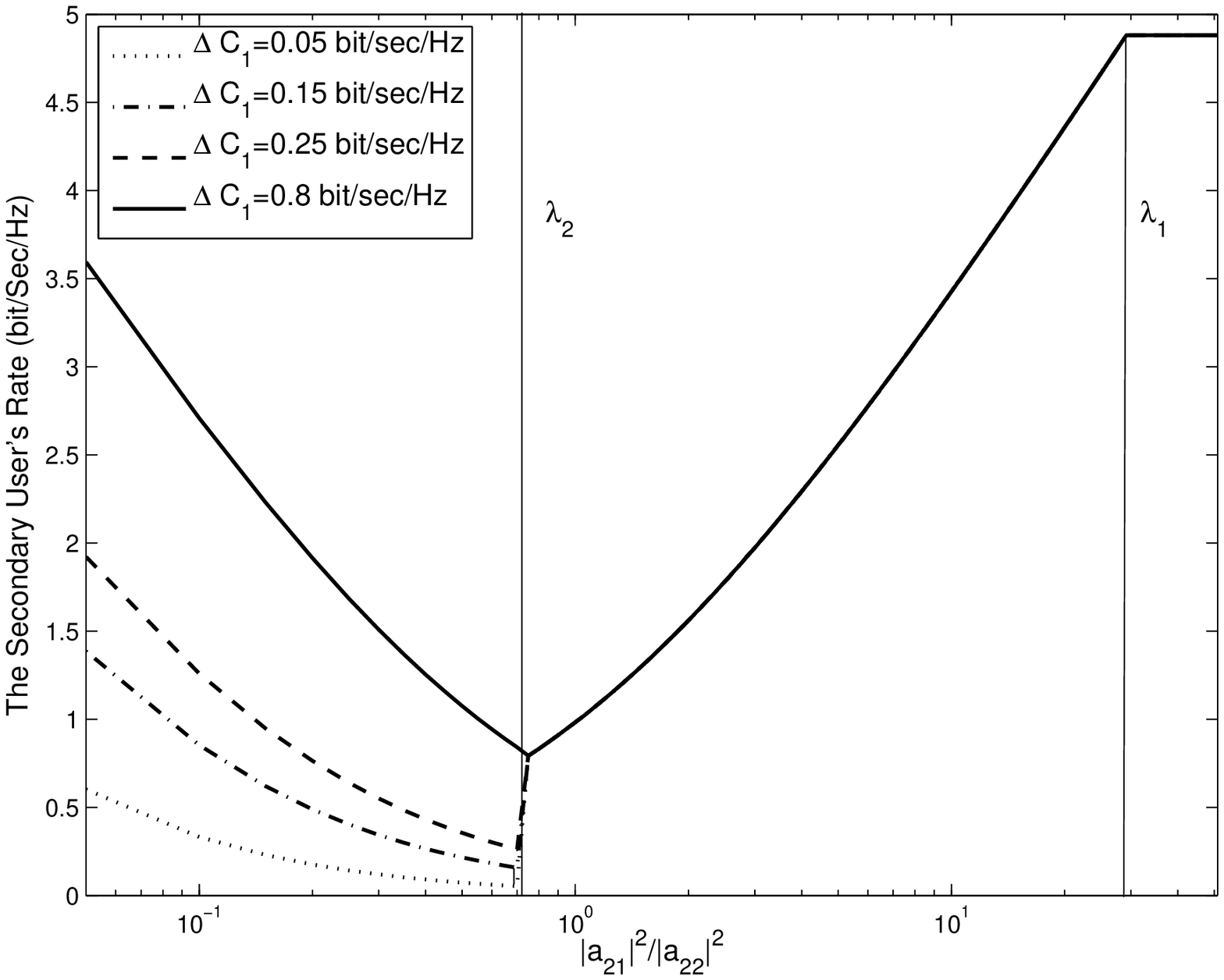, width=1\columnwidth}
\end{center}
\caption{An example of capacity results for the PSMD mode with
full multiuser CQI.}
\label{figure5}
\end{figure}

\begin{figure}[t]
\begin{center}
\epsfig{file=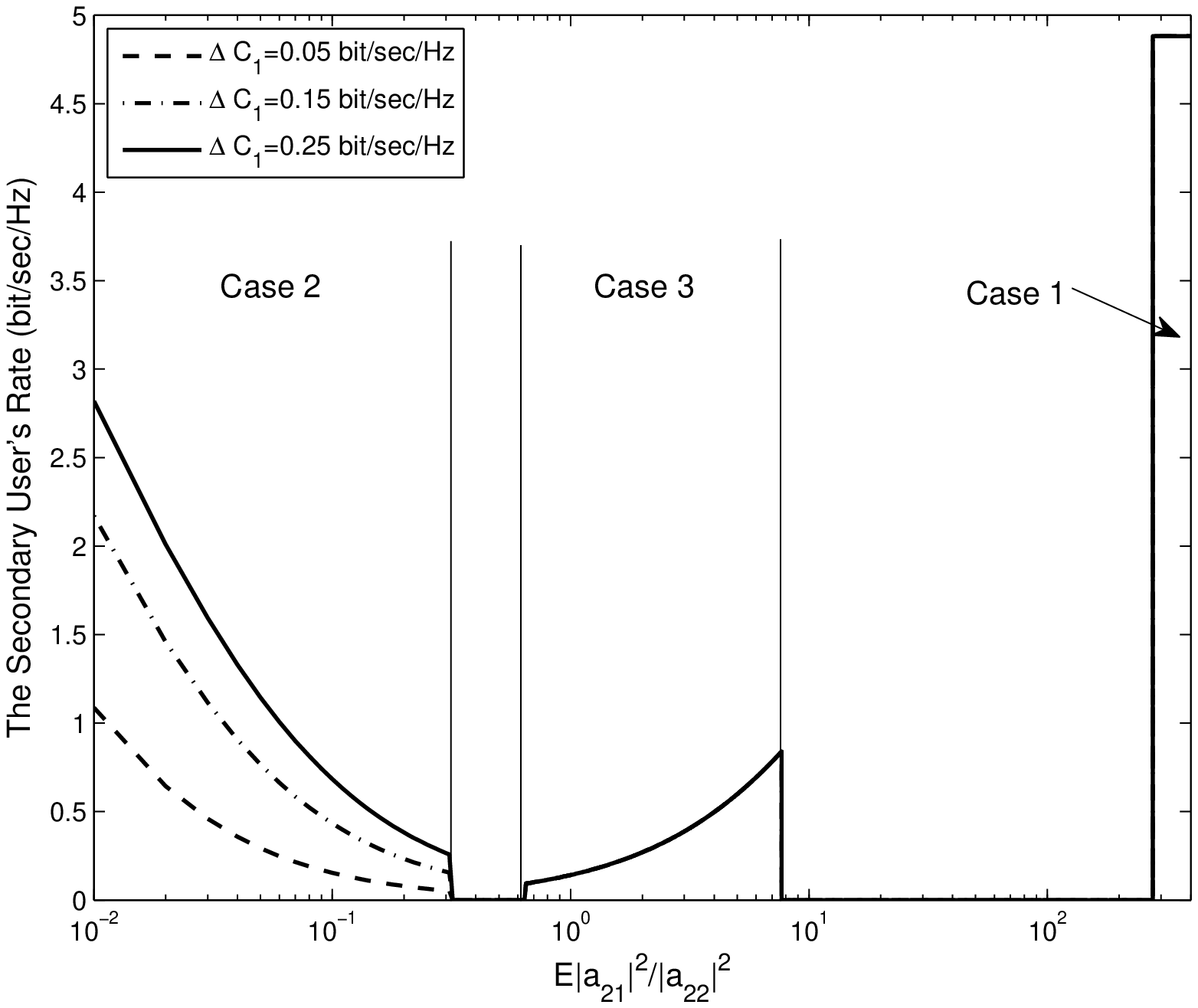, width=1\columnwidth}
\end{center}
\caption{An example of capacity results for the PSMD mode with
partial multiuser CQI.}
\label{figure6}
\end{figure}

Next, we use a visual example to exhibit the performance. The system 
configuration is the same as the setup in Section III-C. For the 
scenario with full multiuser CQI, \Fig{figure5}
shows the secondary user's achievable rate as a function of 
the ratio $|a_{21}^{}|^2/|a_{22}^{}|^2$. Calculation of the achievable 
rate follows the conclusion in Section V-A. For the scenario with partial
multiuser CQI, \Fig{figure6} shows the secondary user's achievable rate 
as a function of the ratio $E(|a_{21}^{}|^2)/|a_{22}^{}|^2$. Calculation 
of the achievable rate follows the results presented in {\em Corollaries 1-4} 
by setting the outage probability $\mathcal{O}_\mathrm{t}=\mathcal{O}_1=10\%$ 
and the probability $\epsilon=90\%$. It is observed that {\em Case 1} will happen
only for the condition $E(|a_{21}^{}|^2)/|a_{22}^{}|^2>300$, which often does not
hold in practice. {\em Case 3} requires the primary user to operate at a SNR 
larger than $15$ dB (see {\em Corollary 3}). However, in this case, the 
secondary user cannot gain more than $1$ bit/sec/Hz at $P_2/N_o=16$ dB. 
Finally, {\em Case 2} shows a comparable performance with the corresponding 
scenario ($|a_{21}^{}|^2/|a_{22}^{}|^2<\lambda_2$) in \Fig{figure5}.

\section{The TSMD Mode}
\Fig{figure2}-(d) depicts the TSMD mode where each user knows the other's
codebook. Then, each user has the chance to decode the other user's
message so as to cancel the interference.

\subsection{Capacity Results with Full Multiuser CQI}
Capacity theorem about two-user GIC channel \cite{D1355} has told us
that the secondary user cannot reliably decode the primary user's message
for the channel condition $|a_{12}^{}|<|a_{11}^{}|$. Hence, for the case of
$|a_{12}^{}|<|a_{11}^{}|$, the TSMD mode reduces to a special example of
the PSMD mode.

For the channel condition $|a_{12}^{}|\geq|a_{11}^{}|$
and $|a_{21}^{}|\geq|a_{22}^{}|$,
the TSMD system becomes a compound multiple-access channel \cite{Sato81}
whose capacity region is given by \Ref{equ03}. In this case, the primary
user does not need to pay capacity penalty, and thus the secondary user's
capacity is
\begin{IEEEeqnarray}{ll}\label{equ47}
C_2&=\min\left(\mathcal{C}\left[\gamma_{21}^{}+\gamma_{11}^{}\right], \mathcal{C}\left[\gamma_{12}^{}+\gamma_{22}^{}\right]\right)
-\mathcal{C}\left[\gamma_{11}^{}\right].
\end{IEEEeqnarray}
The transmit power $P_2$ is limited only by the local power constraint.

For the channel condition $|a_{12}^{}|\geq|a_{11}^{}|$ and
$|a_{21}^{}|<|a_{22}^{}|$, the secondary user can access the primary spectrum
without causing capacity penalty to the primary user. In this case, each user
will decode the other's message for interference cancelation, and thus the
secondary user's transmission rate is \Ref{equ47}. Due to $|a_{21}^{}|<|a_{22}^{}|$,
we can easily justify that \Ref{equ47} equals to \Ref{equ27}. If the primary user deals with the interference as noise, the TSMD mode reduces to the SSMD mode.
Then, the secondary user's transmission rate is \Ref{equ21}, and the transmit
power $P_2$ is limited by \Ref{equ11}. According to {\em Criteria 2\&3},
the secondary user's achievable rate for the channel condition
$|a_{12}^{}|\geq|a_{11}^{}|$ and $|a_{21}^{}|<|a_{22}^{}|$ is
\begin{equation}\label{equ48}
R_2<\max(\Ref{equ21}, \Ref{equ27}).
\end{equation}

\subsection{The UCR Strategy with Partial Multiuser CQI}
It has been shown in Sec. VI-A that the TSMD mode reduces to the
PSMD mode for the channel condition $|a_{12}^{}|<|a_{11}^{}|$. Therefore,
the UCR strategy here is proposed only for the condition
 $|a_{12}^{}|\geq|a_{11}^{}|$.

\subsubsection*{Case 1}
Suppose
\begin{equation}\label{equ49}
\mathrm{Pr}(|a_{21}^{}|^2\geq|a_{22}^{}|^2)>\epsilon
\end{equation}
the secondary user will access the primary spectrum at the
transmission rate
\begin{equation}\label{equ50}
R_2\leq\min\left(\mathcal{C}\left[\frac{P_2\mathcal{L}}{N_o}+\gamma_{11}^{}\right], 
\mathcal{C}\left[\gamma_{12}^{}+\gamma_{22}^{}\right]\right)
-\mathcal{C}\left[\gamma_{11}^{}\right].
\end{equation}
\Ref{equ50} is produced by replacing the term $\gamma_{21}^{}$ in \Ref{equ47}
with $(P_2\mathcal{L})/(N_o)$ where $\mathcal{L}>|a_{22}|^2$.

\subsubsection*{Case 2}
Suppose
\begin{equation}\label{equ51}
\mathrm{Pr}(|a_{21}^{}|^2<|a_{22}^{}|^2)>\epsilon
\end{equation}
the UCR strategy is described as the following steps:

\subsubsection*{Step 1}
Utilize \Ref{equ14} to determine $\max(P_2)$ with respect to a
given capacity penalty $\Delta C_1$;

\subsubsection*{Step 2}
Calculate the following result which is produced by
replacing $P_2$ in (\ref{equ21}) with \Ref{equ14}
\begin{equation}\label{equ52}
C_2^{(21)}=\min\left(\mathcal{C}\left[\gamma_{12}^{}
+\frac{\max(P_2)|a_{_{22}}|^2}{N_o}\right]
-(1-\rho)\mathcal{C}\left[\gamma_{11}^{}\right],
\mathcal{C}\left[\frac{\max(P_2)|a_{_{22}}|^2}{N_o}\right]\right)
\end{equation}

\subsubsection*{Step 3}
Calculate the following result which is produced by
replacing the term $\gamma_{21}^{}$ in \Ref{equ27} with
$(P_2\mathcal{L})/(N_o)$ ($\mathcal{L}<|a_{22}^{}|^2$)
\begin{equation}\label{equ53}
C_{2}^{(27)}=\mathcal{C}\left[\frac{P_2\mathcal{L}}{N_o}
+\gamma_{11}^{}\right]
-\mathcal{C}\left[\gamma_{11}^{}\right].
\end{equation}

\subsubsection*{Step 4}
Determine the secondary user's transmission rate via
$R_2\leq\max(C_{2}^{(21)}, C_{2}^{(27)})$.

\subsection{Numerical Example}
Considering $|a_{21}^{}|$ to be Rayleigh distributed, we derive the
following results for {\em Case 1} and {\em Case 2}, respectively.
\begin{cor}
A sufficient condition for {\em Case 1} to happen is
\begin{equation}\label{equ54}
E(|a_{21}^{}|^2)\geq\frac{|a_{22}^{}|^2}{\ln(1/\epsilon)},~\mathrm{and}~
E(|a_{21}^{}|^2)\geq\frac{\mathcal{L}}{\ln(1/\epsilon)}
\end{equation}
\end{cor}
\begin{IEEEproof}
\Ref{equ54} can be straightforwardly obtained through calculation of
\Ref{equ49} and $\mathrm{Pr}(\mathcal{L}\leq|a_{21}^{}|^2)\leq\epsilon$.
\end{IEEEproof}

\begin{cor}
A sufficient condition for {\em Case 2} to happen is
\begin{equation}\label{equ55}
E(|a_{21}^{}|^2)\leq\frac{|a_{22}^{}|^2}{\ln(1/(1-\epsilon))},~\mathrm{and}~
E(|a_{21}^{}|^2)\geq\frac{\mathcal{L}}{\ln(1/\epsilon)}
\end{equation}
\end{cor}
\begin{IEEEproof}
\Ref{equ55} can be straightforwardly obtained through calculation of
\Ref{equ51} and $\mathrm{Pr}(\mathcal{L}\leq|a_{21}^{}|^2)\leq\epsilon$.
\end{IEEEproof}

\begin{figure}[t]
\begin{center}
\epsfig{file=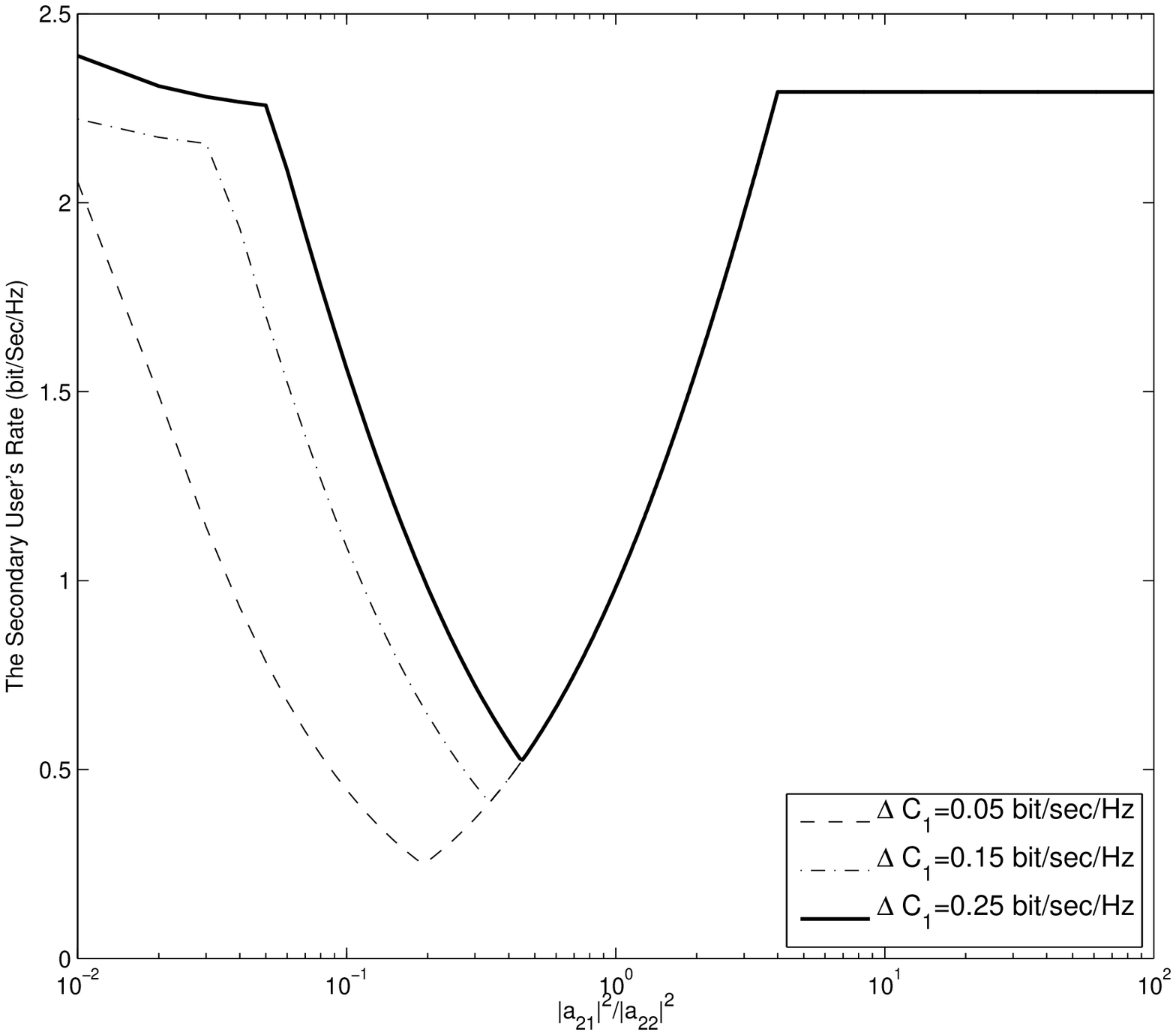, width=1\columnwidth}
\end{center}
\caption{An example of capacity results for the TSMD mode with
full multiuser CQI.}
\label{figure7}
\end{figure}

\begin{figure}[t]
\begin{center}
\epsfig{file=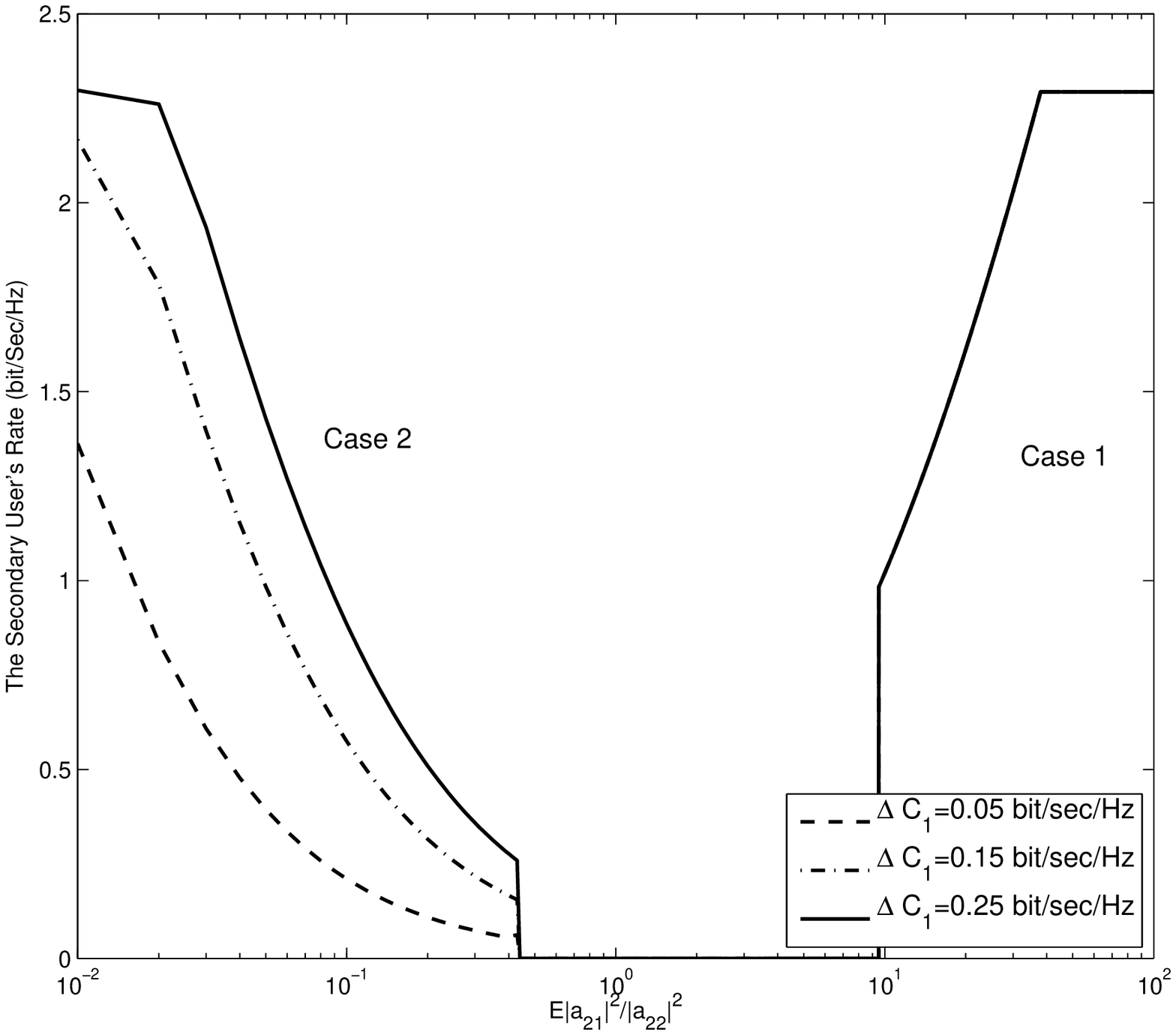, width=1\columnwidth}
\end{center}
\caption{An example of capacity results for the TSMD mode with
partial multiuser CQI.}
\label{figure8}
\end{figure}

Figs. \ref{figure7}-\ref{figure8} show a visual example for scenarios with 
full or partial multiuser CQI, respectively. The system configuration 
is almost the same as the setup in Section III-C, but we set 
$|a_{12}^{}|^2=4$ to fulfill the condition $|a_{12}^{}|>|a_{11}^{}|$.  
For the scenario with partial multiuser CQI, we set 
$\mathcal{O}_\mathrm{t}=10\%$ and $\epsilon=90\%$ as an example. 
It is observed that {\em Cases 1-2} in \Fig{figure8} offers comparable 
performance with the corresponding scenario in \Fig{figure7}. 

\section{Conclusion}
In this paper, we have investigated two-user Gaussian UCR systems 
by assuming the availability of full multiuser CQI or partial multiuser 
CQI. Provided full multiuser CQI, we have studied the fundamental 
relationship between the secondary user's achievable rate $C_2$ and 
capacity penalty to the primary user $\Delta C_1$ in four carefully classified 
UCR modes. For the scenario with partial multiuser CQI, we first 
established a new physical-layer model through exploitation of the 
location-aided approach. Then, new spectrum access and power allocation 
strategies have been investigated in terms of $C_2$, $\Delta C_1$, and 
capacity outage probability. Numerical examples are provided to show 
the performance of the UCR with full multiuser CQI and the proposed 
approach with partial multiuser CQI.

\section*{Appendix}
\subsubsection*{Proof of Corollary 3}
For the Rayleigh distribution, we can calculate
\begin{IEEEeqnarray}{ll}\label{app1-1}
&\mathrm{Pr}\left(\lambda_2
\leq\frac{|a_{_{21}}|^2}{|a_{_{22}}|^2}\leq
\lambda_1\right)\nonumber
\\&\quad\quad=\exp\left(-\frac{\gamma_{22}^{}\lambda_2}
{\bar{\gamma}_{21}^{}}\right)
-\exp\left(-\frac{\gamma_{22}^{}\lambda_1}
{\bar{\gamma}_{21}^{}}\right)
\\&\quad\quad\triangleq f(\bar{\gamma}_{21}^{})\nonumber
\end{IEEEeqnarray}
Using the first derivative of $f(\bar{\gamma}_{21}^{})$ with
respect to $\bar{\gamma}_{21}^{}$, we can find that
$f(\bar{\gamma}_{21}^{})$ is an increasing function of
$\bar{\gamma}_{21}^{}$ for the condition
\begin{equation}\label{app1-2}
\bar{\gamma}_{21}^{}\leq
\frac{\gamma_{22}^{}(\lambda_1-\lambda_2)}
{\ln\left(\lambda_1/\lambda_2\right)}
\end{equation}
and otherwise a decreasing function. Hence, we have
\begin{IEEEeqnarray}{ll}\label{app1-3}
\max\left(f(\bar{\gamma}_{21}^{})\right)
&=f\left(\bar{\gamma}_{21}^{}=\frac{\gamma_{22}^{}(\lambda_1-\lambda_2)}
{\ln\left(\lambda_1/\lambda_2\right)}\right)\nonumber
\\&=\exp\left(-\frac{\ln(\gamma_{11}^{}+1)}{\gamma_{11}^{}}\right)
\left(\frac{\gamma_{11}^{}}{\gamma_{11}^{}+1}\right)
\end{IEEEeqnarray}
A necessary condition for \Ref{equ33} to be satisfied is
$\max\left(f(\bar{\gamma}_{21}^{})\right)>\epsilon$. Due to
$(\gamma_{11}^{})/(\gamma_{11}^{}+1)<1$, it is necessary to
have the following condition to be satisfied
\begin{equation}\label{app1-4}
\exp\left(-\frac{\ln(\gamma_{11}^{}+1)}{\gamma_{11}^{}}\right)>\epsilon.
\end{equation}
Solving this inequality leads to \Ref{equ40}.

\section*{Acknowledgment}
The authors would like to thank the editor, Dr. Ronald Raulefs, and anonymous reviewers for their extremely constructive and supportive comments.

\bibliography{nabib,mybib,nabib1}
\bibliographystyle{ieeepes}

\newpage

\end{document}